%% file: vanBelle.tex
\documentclass[12pt,preprint]{aastex}

\slugcomment{Accepted for the September 2002 Astronomical Journal}

\shorttitle{Angular Sizes of Mira Variables}
\shortauthors{van Belle, Thompson, \& Creech-Eakman}

\begin{document}


\title{Angular Size Measurements Of Mira Variable Stars At 2.2 $\mu$m. II}


\author{G. T. van Belle\altaffilmark{1}}
\author{R. R. Thompson}
\author{M. J. Creech-Eakman\altaffilmark{2}}
\affil{Jet Propulsion Laboratory, California Institute of Technology, Pasadena, CA 91109}


\altaffiltext{1}{Email address: gerard@huey.jpl.nasa.gov.}
\altaffiltext{2}{Caltech/JPL Postdoctoral Scholar.}


\begin{abstract}
We present angular size measurements of 22 oxygen-rich Mira
variable stars.  These data are part of a long term observational
program using the Infrared Optical Telescope Array (IOTA) to
characterize the observable behavior of these stars. Complementing
the infrared angular size measurements, values for variable star
phase, spectral type, bolometric flux and distance were
established for stars in the sample; flux and distance led to
values for effective temperature ($T_{EFF}$), and linear radius,
respectively. Additionally, values for the $K-[12]$ color excess
were established for these stars, which is indicative of dusty
mass loss. Stars with higher color excess are shown to be
systematically 120 $R_\odot$ larger than their low color excess
counterparts, regardless of period.
This analysis appears to present a solution to a long-standing question
presented by the evidence that some Mira angular diameters are
indicative of first overtone pulsation, while other
diameters are more consistent with fundamental pulsation.
A simple examination of the
resultant sizes of these stars in the context of pulsation mode
is consistent with at least some of these objects pulsating in
the fundamental mode.
\end{abstract}


\keywords{stars: variables: Mira, infrared: stars, stars:
fundamental parameters, stars: late-type,
techniques:interferometric}


\section{Introduction}

The recent development of interferometric methods at optical and
infrared wavelengths has provided the astronomical community with
more than an order of magnitude increase in spatial resolution
over direct imaging techniques.  Using the Infrared Optical
Telescope Array (IOTA, see \citet{ca94} and \citet{dy95})
and the Palomar Testbed Interferometer (PTI, see \citet{co99}), we have
been conducting a program of high resolution, K band observations
of Mira variable stars.  Previous IOTA results have been presented for
both oxygen-rich Miras \citep{va96}, and carbon \& S-type Miras
and non-Mira S-type stars \citep{va97}.

Mira variables figure prominently in our observing strategy, since
these stars are large and bright at infrared wavelengths.  Using
crude estimates of surface temperatures and the observed total
fluxes it is estimated that more than 100 such stars have black
body angular diameters in excess of 2 milliarcseconds (mas),
easily resolvable targets for the current generation of Michelson
interferometers.  The 22 stars reported herein represent a
significant collection of angular sizes to be added to the
literature, roughly doubling the number of sizes available.  This
large sample allows for a more detailed analysis of the behavior
of these stars.

Recent investigations of Mira variables have attempted to resolve
important questions regarding the pulsation mode, mass loss and
evolution of these stars.  The pulsation mode remains a
currently-unresolved issue; the promise of high-resolution
interferometric methods to firmly resolve this question is
hampered by conflicts between distance-determination methods.
Initial studies of these stars in the
visible \citep{ha95,la97,bu98,yo00},
and
infrared \citep{va96,va97,pe99,ho02,th02b},
have yet to provide conclusive evidence
in favor of fundamental versus first overtone pulsation mode,
although a deeper understanding of these complex and dynamic
stellar atmospheres is beginning to emerge as a direct result of
these investigations.

Studies attempting to derive distances to Miras, independent of
assumptions about pulsation mode, fail to reach agreement on the
subject.  \citet{wh00a}, utilizing parallax data from the
Hipparcos satellite to derive a K-band period-luminosity relation,
which, in conjunction with Mira angular size data and published
model atmospheres, they find to indicate a common first overtone
mode for most Miras. The resultant LMC distance modulus Whitelock
and Feast derive is $18.64\pm0.14$, which is consistent with
similar determinations utilizing Cepheids \citep{gr00} but in
conflict with recent LMC distances as indicated by eclipsing
binaries \citep{ne00, ri00, gr01} that favor smaller distance
modulus values of $18.46\pm0.06$ or less. In contrast to that
investigation, \citet{al97}, utilizing TiO and VO band strengths,
arrive at radii and temperatures that favor fundamental
pulsations.  This finding is in agreement with the group of
\citet{wo99, wo00}, which examines Mira period ratios found in
MACHO data to argue strongly in favor of fundamental mode.
Depending upon initial assumptions, models supporting both
fundamental mode pulsation \citep{wi79,hi79,wo90} and first
overtone pulsation \citep{wo74, tu79, pe90} have been
constructed, in certain cases, the pulsation mode itself is a free
parameter \citep{ho98}.

It is by means of direct observation of the angular sizes of Mira
variables that we may be able to provide unique insight into these
questions.
Additional information in the form of bolometric flux
estimates will yield further information, such as effective
temperature, which remains a poorly established quantity for this
class of stars. Also, our angular size measurements and derived
quantities have implications regarding the nature of mass loss and
evolution among Mira variables.

Carbon-rich, oxygen-rich and S-type Miras were all observed at
IOTA and the Palomar Testbed Interferometer (PTI)
as a part of our ongoing high resolution program; in this
paper we present only the observations of the oxygen-rich variety observed at IOTA.
Observations at IOTA using the related FLUOR experiment
are not considered in this article as part of the data set \citep{pe99},
although they are in agreement with our conclusions, particularly
that of fundamental mode pulsation for Miras \citep{ho02}.
Operations at IOTA that produced these results are discussed in
\S 2, detailing source selection and observation.  In \S 3 the
procedures used in establishing the stellar parameters for the
stars observed are discussed; the parameters include phase,
spectral type, bolometric flux, angular size, effective
temperature and linear radius.  These parameters are in turn
examined for significant interrelationships in the discussion of
\S 4.

\section{Observations}

The data reported in this paper were obtained in the $K$ band
($\lambda$ = 2.2 $\mu$m, $\Delta\lambda$ = 0.4 $\mu$m) at IOTA,
using the 21 and 38 meter baselines.  Use of IOTA at 2.2 $\mu$m to
observe Mira variables offers three advantages: First, effects of
interstellar reddening are minimized, relative to the visible
($A_K = 0.11 A_V$; \citet{ma90}); Second, the effects of
circumstellar emission are minimized shortward of 10 $\mu$m
\citep{ro83}, and; Third, the $K$ band apparent
uniform-disk diameter of Mira variables has in the past been
expected to be close to
the $\tau=1$ Rosseland mean photospheric diameter (see the discussions in
\S \ref{tad} and \S\ref{discussion}). The interferometer, detectors and data reduction procedures
have been described more fully by \citet{ca94} and \citet{dy95}.
As was previously reported in these papers, starlight collected by
the two 0.45 m telescopes is combined on a beam splitter and
detected by two single element InSb detectors, resulting in two
complementary interference signals.  The optical path delay is
mechanically driven through the white light fringe position to
produce an interferogram with fringes at a frequency of 100 Hz.
Subsequent data processing locates the fringes in the raw data and
filters out the low and high frequency noise with a square filter
50 Hz in width.  Recent software and hardware upgrades to the
computers driving the interferometer and collecting the data have
resulted in a improved data collection rate to 1500-2000 fringe
packets per night.  On the best nights we can observe 20 science
sources and an equal number of calibrators.

Observations of target objects are alternated with observations of
unresolved calibration sources to characterize slight changes in
interferometer response (no more
than a few percent), due to both seeing and instrumental
variations.  Raw visibilities are determined from the amplitude of
the interferogram at the white light fringe position, normalized
by the incoherent flux from the star.  An estimate of the noise is
obtained from a similar measurement made in the data
outside the region of coherence; the noise estimate is used in
obtaining a weighted average of the visibility data, which is
typically taken in sets of 50 interferograms.  The raw
visibilities of the target objects are then calibrated by dividing
them by the measured visibilities of the calibration sources,
using the calibration sources as samples of the interferometer's
point response.  Calibration sources were selected from $V$ band
data available in The Bright Star Catalog, 4th Revised Edition
\citep{ho82} and $K$ band data in the Catalog of Infrared
Observations \citep{ge93}, based upon angular sizes calculated
from estimates of bolometric flux and effective temperature;
calibration source visibility was selected to be at least 90\% and
ideally greater than 95\%, limiting the effect of errors in
calibrator visibility to a level substantially below measurement
error.

Mira variables observed for this paper were selected based upon a
number of criteria.  Stars needed to be bright enough in $V$ and $K$
to be detected by both the star trackers and the InSb detectors;
the current limits of the IOTA interferometer dictate $V < 8.0$ mag
and $K < 5$ mag (though for observations at all airmasses and seeing
conditions, we require $K < 2.5$ mag).  The Mira variables needed
to be at a declination accessible to the mechanical delay
available for a given evening.  This is because the difference in
delay between the two apertures, which can range from -30 m to +20
m, depends upon source declination and hour angle.  Since only 4.6
m of this range is accessible at any time, observing is
constrained to a specific declination bin about $10 \deg$ wide on
any given night.  The stars also needed to be of sufficient
estimated angular size to be resolved by IOTA.  Mira phase was not
a factor in target selection; hence, our targets represent Miras
at a variety of phases, from visible light maximum to minimum.

Twenty-two oxygen-rich Mira variables were observed at IOTA during
four observing runs in March, May/June, October of 1996, and July
of 1997.  The visibility data for the two detector channels have
been averaged and are listed in Table 1, along with the date of
the observation, the interferometer projected baseline, the
stellar phase and the derived uniform disk angular size; the
latter two points are discussed further in \S 3.  In our experience
with the IOTA interferometer, \citet{dy96} has demonstrated that
the night-to-night RMS fluctuations in visibility data generally
exceed the weighted statistical error from each set of
interferograms; we have characterized these fluctuations and use
the empirical formula $\sigma_V = \pm 0.051/\sqrt{N}$ (number of
nights) to assign the ``external'' error.  The interested reader
should see \citet{dy96} for a more complete discussion.  Finally,
visibility data were fit to uniform disk models to obtain an
initial angular size $\theta_{UD}$.  These uniform disk diameters
and their estimated errors, derived from the uncertainty in the
visibilities, are also listed in Table 1.

We note that typically a single point was utilized in calculating
the uniform disk diameter $\theta_{UD}$.  For the stars in our
sample, the visibility data were all at spatial frequencies, $x$,
shortward of the first zero of the uniform disk model, $\left|2
J_1(x) / x\right|$, where $x=\pi\theta_{UD}B / \lambda$.
\citet{ha95} noted that the uniform disk model was
not a particularly good model for visible-light data for Mira
variables; rather, the data were a better fit to a simple
Gaussian.  Although we do not currently have multiple spatial
frequency data for any Mira variables, our naive expectation
as followed in \citet{va96, va97} was that the
departures from a uniform disk model will not be as great at 2.2
$\mu$m as it is at visible wavelengths.
This expectation appears to be borne out by comparisons of
the data for R Aqr in \citet{va96} with the observations
of \citet{tu00}, although as we shall see in \S \ref{discussion},
this may not be entirely accurate.
Thus, initially we will assume that to first order, a uniform disk
model will also fit the Mira data; a slight correction to the
derived angular sizes to account for this assumption will be
discussed in \S\ref{tad}.
In this case, a single spatial frequency point
will uniquely and precisely determine the angular diameters for
visibilities in the approximate range $0.25 \geq V \geq 0.75$.
As we shall see in \S\ref{discussion},
there is evidence this approach does not properly account for
the deviation of a uniform disk size from the Rosseland mean diameter.
Properly examining
this point may only be addressed by detailed multiple
spatial frequency observations of the visibility curves.




\section{Stellar Parameters}

\subsection{Phase} \label{phase}

The phases of the Mira variables observed were established by
means of two sources, following the procedure outlined in \citet{va96, va97}.
Periods were initially obtained from The
General Catalog of Variable Stars, 4th Edition (GCVS,
\citet{kh88}).  However, since the zero phase date in the GCVS at
the epoch of the observations was no less than 11 cycles old for
our sample stars, visual brightness data available from the
Association Francaise des Observateurs d'Etoiles Variables (AFOEV)
was utilized in estimating a recent zero phase date
\citep{sc98}.

As an additional cross-check, Fourier analysis (as discussed in
\citet{sc82} and \citet{ho86}) of the AFOEV data also provided
period information, but using light curve data which was more
recent than that found in the GCVS.  The periods from the GCVS and
the AFOEV analysis agreed at the 1\% level, corresponding to an
average difference in period of $1\pm4$ days. With the agreement
in periods, the zero phase estimate was the larger uncertainty in
phase determination, although this uncertainty was still small,
averaging 6d.  Periods, determined from Fourier analysis of the
AFOEV data, and phases for each of the Mira variables are
presented in Table 1.

\subsection{Spectral Type \& Bolometric Flux}

Bolometric fluxes ($F_{BOL}$) of the Mira variable stars were
estimated from a relationship between $F_{BOL}$ and 2.2 $\mu$m
flux ($F_K$), as established by \citet{dy74}.  In order to obtain
bolometric fluxes, $K$ magnitudes were first estimated from the
incoherent (off-fringe) flux levels present in the IOTA data.  We obtained our
standard star photometric calibrations using the $K$ band
measurements found in the Two Micron Sky Survey \citep{ne69} for
our non-variable point-response calibration sources.    Cross-calibrator comparison of the published
versus measured values for $m_K$ indicates the IOTA photometry
is consistent with the previous measures, with uncertainty
being dominated by measurement scatter.
No airmass
corrections were applied since the calibrators were observed at
nearly identical airmasses as the Mira variables.  In all cases
the bolometric fluxes were obtained from the absolute $K$ fluxes
through the observed relation
\begin{equation}\label{eqn32}
\log (F_K/F_{BOL}) = 0.017\times(V-K)-0.74
\end{equation}
which is the mean relationship derived from \citet{dy74} for
spectral types M5-M10.  A 15\% error bar was assigned to the
resultant $F_{BOL}$ values, which is consistent with more detailed
$F_{BOL}$ estimations done by \citet{wh00a}. We note that the $\log
(F_K/F_{BOL})$ - spectral type relationship also has a firm
theoretical basis and may be seen in the ``infrared flux method''
calculations carried out by \citet{bl94}. No reddening corrections
were applied to obtain the bolometric fluxes.  These were deemed
unnecessary since the typical magnitude of the corrections will be
$\Delta m_K < 0.10$ mag (calculated from the empirical reddening determination
of Mathis, 1990, and from the $A_V$ values given for local Miras
in Whitelock et al. 2000), is less than the RMS K band error,
$\Delta m_K = 0.15$ mag.

In contrast to our previous paper on this subject \citep{va96},
spectral types were {\it not} inferred from the mean observation data of
\citet{lo71} or \citet{lo72}. We note instead that over the range
of spectral types in question (M5-M9.8), the $\log(F_K/F_{BOL})$
vs. $V-K$ relationship is nearly flat, making the determination of
$F_{BOL}$ robust despite possible errors in $V-K$ color. An uncertainty
of one full magnitude in $V-K$ results in only a 4\% difference in
the determined $F_{BOL}$; we shall see in \S\ref{temperature} that this results
in a negligible difference in the determined effective
temperature. The errors in the $K$ magnitudes are the standard
deviations of the individual measurements on a given night.  From
the observed scatter in the $F_K/F_{BOL}$ relationship we estimate
an rms error of $\pm13$\% in $F_{BOL}$ from the use of the $K$
magnitude; we estimate a further uncertainty of $\pm5$\% in the
absolute calibration \citep{bl94}. The estimated error for
$F_{BOL}$ in Table 2 is the quadrature sum of these contributions
of 15\%.

\subsection{True Angular Diameter}\label{tad}

In order to estimate effective temperatures, the uniform disk
diameters in Table 1 needed to be converted to stellar diameters
corresponding to the non-uniform extended atmospheres of the Mira
variables.  We used the model Mira
atmospheres discussed in \citet{ho98}[HSW98].  We note that the
HSW98 models did not account for the time for both shock
compressed material and material expanding between shocks to
return to radiative equilibrium.  These regions, with $T >
T_{RadEq}$ and $T < T_{RadEq}$, respectively, can alter the
brightness distribution profile and consequently alter the `true'
angular sizes derived from the uniform disk (UD) diameters.
Rather, HSW98 follows \citet{be96}
in the semi-empirical adoption of an equilibrium temperature just behind the
shock front.
Dynamical atmosphere calculations have the potential to resolve
these concerns \citep{bo88,bo91}; however, center-to-limb
brightness profiles are not yet available for such calculations.
The missing physics in the models has the potential to make for
poor agreement between angular sizes derived at different
wavelength bands.

Noting these concerns, here we shall use the
HSW98 models as a sufficient expectation of the intensity
distribution across the disk of a Mira variable to proceed with
our analysis.  Fortunately, as
illustrated in HSW98, the K band is a particularly forgiving
bandpass in which to work, and the differences between UD
diameters and the Rosseland mean radiating surface is potentially small.
Examining all of the HSW98 models, four of the six models (series Z, D, E and O) were
seen as most representative of the parameters that matched the
Miras observed; specifically, examining the bolometric flux
variations of the local Miras as seen in \citet{wh00a}, the average
$\Delta m_{BOL}=0.73\pm0.25$, which compares well with the
bolometric flux variations seen in those four models in HSW98. For
those models, the rms difference between a uniform disk fit and
the Rosseland mean diameter is $1.00\pm0.04$, which we shall use
as a scaling factor from which Rosseland angular sizes can be
derived from K band UD diameters.
Within the context of a unit scaling factor, the main quantitative impact
upon the Rosseland mean diameter from the UD diameter is
the slight increase in error due to the uncertainty in scaling factor.
A scaling factor equal to unity is consistent with
expectation that near-infrared uniform disk diameters should be
reasonable as direct indicators of the true photospheric diameters
of Mira variables (\citet{wi86}), although contrasting with the possibility
that extended atmospheric constituents might affect the apparent
stellar size (\citet{be01}).  As we shall see in \S\ref{discussion}, this
appears to be more than a mere possibility.

In our previous article on Miras observed at IOTA \citep{va96}, we
utilized a phase-dependent scaling as derived from \citet{sc87},
ranging from 0.98 at maximum light to 1.11 at minimum light.  For
this article, we decided to utilize a constant scaling in order to
increase our sensitivity to true size and temperature variations,
rather than variations derived from varying scaling factors.  Our
UD diameters from \citet{va96} were re-scaled using the $1.00\pm0.04$ value
above as well.

We also note that some consideration was given to the possibility
of departures from spherical symmetry in these variable stars.  As
has been observed in visible light observations of Mira variables
\citep{ka91,ha92,wi92, we96, tu99}, these stars can be considerably elongated,
possessing up to a 20\% difference between semi-major and
semi-minor axes, although this appears to diminish at 1.0 $\mu$m \citep{ho00}.
Similar observations at 2.2 $\mu$m have indicated that
this elongation is potentially present in the near-IR as well,
with asymmetries of up to 25\%
\citep{tu00,th02}.
Such K band asymmetries potentially could be explained in terms
of high-layer water or other molecular contamination
seen to be present in models at certain combinations of
parameters and phase cycle \citep{be01,sc01},
and are consistent with recent observations of Miras using
0.1 $\mu$m bands across the K band window \citep{th02b}.
The reader should be aware of the potential for
this effect to introduce spread in our sample of angular sizes,
although it is our expectation that its magnitude will be no greater than
the errors in the data set.

\subsection{Effective Temperature}\label{temperature}

The stellar effective temperature, $T_{EFF}$, is defined in terms
of the star's luminosity and radius by $L = 4\pi \sigma R^2
T_{EFF}^4$. Rewriting this equation in terms of angular diameter
and bolometric flux, a value of $T_{EFF}$ was calculated from the
flux and Rosseland diameter using
\begin{equation}\label{tempeqn}
T_{EFF} = 2341 \times {\left({F_{BOL} \over
\theta_R^2}\right)}^{1/4}
\end{equation}
the units of $F_{BOL}$ are $10^{-8}$ erg cm$^{-2}$s$^{-1}$, and $\theta_R$
is in mas. The error in $T_{EFF}$ is calculated from the usual
propagation of errors applied to equation \ref{tempeqn}.  The measured
$T_{EFF}$'s are given in column 8 of Table 2, and are found to
fall in the range between 2000K and 3250K.

\subsection{Linear Radius}\label{linearRadius}

In order to establish a linear radius from the angular size data
presented herein, a distance estimate to the observed Miras needs
to be established.  Unfortunately, such an estimate can be rather
contentious, particularly with the implications upon pulsation
mode as can be seen in \citet{ha95} and \citet{va96}. Herein we
shall consider the infrared period-luminosity relationship of
\citet{fe89} as considered by \citet{wi00}.  In this review, the
relationship is seen to fall in two roughly linear relationships
for short and long period Miras, which can be expressed as a
function of absolute $K$ magnitude:
\begin{eqnarray}
M_K = -3.66 \times \log(P)+1.42 & \textrm{ for P $>$ 400}\\
M_K = -6.94 \times \log(P)+10.0 & \textrm{ for P $<$ 400}
\end{eqnarray}
The above period-luminosity
relationship carries the implicit assumption that all Miras are
pulsating in the same mode; given the variability of Miras at $K$ and other uncertainties in
this relationship, we shall assume that distances derived from
it have 25\% errors.
The results which will be established in \S 4 were also considered in
light of the $K$ band period-luminosity relationship
given in \citet{wh00a} ($M_K=-3.47 \times \log(P) + 0.84$ for all periods),
which favored an overtone pulsation mode, and resulted in only marginal
quantitative $(\approx 4\%)$ and no qualitative differences in \S 4.
As such, we do not feel that pulsation mode assumptions that might be inherent in
the selected period-luminosity relationship affect our conclusions.

Due to the large standard errors in the Hipparcos Mira data set, we will
not consider distance metrics derived from the parallaxes for
these stars in the otherwise excellent database (eg.
\citet{va97b},\citet{wh00a}). The release of the Hipparcos results
\citep{pe97} has provided the community with a wealth of distance
data on a variety of stars. For Miras, however the results were
rather unimpressive: the parallaxes showed a great deal of scatter
and the standard errors were large (catalog field H16), on average
$2.2\pm3.0$ mas for the stars presented in this study, which is
consistent with the average standard error of $2.6\pm2.5$ mas
found for the 172 M type Miras found in Table 1 of \citet{wh00a}.

There are two potential sources for these effect in the Hipparcos
parallax data: First, a 10 mas object (typical of this study) with
an intrinsic size of roughly 350 $R_{\odot}$ \citep{va97} has a
parallax of approximately 3 mas: the angular diameter of a Mira
variable is about three times its parallax, with the standard
error on a typical parallax being equal to the measurement itself.
Determination of a varying visible light photocenter shift, the
bandpass of Hipparcos operation, could easily be complicated by
this disparity. Second, any spots on the surface of these highly
evolved objects could also affect the apparent position of the
photocenter \citep{va97b}, although the existence of such spots on Miras
has not yet been established empirically. We do not expect this effect to be a
function of the typically large distances to the Miras, relative
to the majority of stars in the Hipparcos data set. A similar
examination of the similarly distant 322 northern hemisphere
single (multiplicity field H59 and spectral type both indicating no
companions) supergiants found in the Hipparcos dataset indicates a
contrasting average standard error of $0.90\pm0.30$ mas; this
average standard error appears to be independent of object
parallax. Furthermore, the possibility of Lutz-Kelker
bias within any given Hipparcos sample of stars has been
empirically established \citep{ou98} and should be carefully
considered.

We have detailed our evaluation of the Hipparcos data set at
length in this section, given our otherwise positive experience
with the catalog's far-reaching utility.  As such, we did not wish to dismiss the
Hipparcos Mira data lightly, but did so only after careful
consideration of the data.


\subsection{Re-Analysis of R Cas}

One star from \citet{va96} that bears
re-examining is R Cas.  In our previous manuscript, a value of
$\theta=13.55\pm0.95$ mas was derived from the measured visibility
$V=0.1259\pm0.0360$ from 4/5 Oct 1995. However, for visibilities
below $V=0.132$, the visibility function becomes non-monotonic and
has multiple solutions.  Neglected in \citet{va96} were the
other angular sizes possible with this visibility: 19.53 and 22.03
mas.  The error envelope about the measured visibility encloses
the peak of the entire first outlying lobe; hence the error
envelope for these angular sizes ranges from 17.90 to 24.16 mas.

Based upon visible light angular size measurements of this
object \citep{ha95, ho00}
that are in the 18 to 36 mas range, the angular size solution most
consistent with our previous $V$ measurement is 22.03 mas.  We note,
though, that for the visibility measurements outside of the
central lobe of the visibility function, the influence of limb
darkening and spotting becomes greater and reduces the effectiveness
of utilizing single visibility measurements to establish stellar
angular sizes, as well illustrated in HSW98.

\section{Discussion: Size \& Pulsation Mode}\label{discussion}

Theoretical models for many Mira variables predict their effective temperatures
will lie at approximately 3,000 to 3,100K, which is somewhat higher than the
values seen in Table 2 (cf. Hoffman et al. 1998, Willson 2000).
One possible explanation for this discrepancy is a systematic
overestimation of the apparent sizes of the variable stars by the interferometer.
Absorption of the central star's flux by molecules will have the effect
of contaminating the observed angular sizes and will lead to a 2.2 $\mu$m
uniform disk size that is markedly larger than the Rosseland mean
diameter.
A few of the models found in HSW98 show `wings'
in the center-to-limb brightness profiles
mainly caused by water molecules \citep{be01, ja02}
that are consistent with this
hypothesis, as do all of the dusty Mira models in \citet{be01}.
As discussed in \S \ref{tad}, conversion of the observed interferometric visibility
to the desired Rosseland $\tau=1$ radius can run afoul of these `wings',
causing the inferred size to be dramatically underestimated, as seen in
\citet{be01} and \citet{sc01}.

To test this hypothesis, we can examine
$K-[12]$ colors for the Miras observed at IOTA. Twelve micron
magnitudes are readily available for these stars, as defined
in \citet{hi95}, using flux values for the observed Miras found in
the Infrared Astronomical Satellite (IRAS) Point Source Catalog
(PSC, 1987). $K-[12]$ is expected to be a reasonable indicator of recent
dusty mass loss (\citet{le96, be90}).  Given our interest in establishing
gross trends in a relative sense between stars of differing $K-[12]$
values, we did not apply point source corrections to the IRAS 12$\mu$m
data \citep{be88}, nor did we attempt to quantify 12$\mu$m variability
for these sources, which is potentially present in any Mira variable
data set \citep{li96,cr97}, but should in this
context only add scatter to the 12$\mu$m values.

K band interferometric observations are relatively insensitive
to dusty mass loss (though not completely so, as evidenced
with VY CMa in \citet{mo99}); however, it is entirely
plausible that dusty
mass loss will be accompanied by a significant increase in the
molecular component of circumstellar material. Significant
amounts of absorption by molecules in a circumstellar shell
above the Rosseland mean radiating surface will have the effect of
increasing the derived size of a interferometrically observed Mira variable.
Taking a black body
at the expected temperature of 3,000K, the predicted $K-[12]$
color will be 0.92.  By comparison to the observed colors for the
individual Miras, we may calculate a $K-[12]$ excess that will be
indicative of stellar mass loss. Given the $K$ band variability
of these objects, we assigned an average error of $\pm0.33$ to our derived
$K-[12]$ excess values,
which is consistent with the observed K band variability of these
objects \citep{wh00b}.
These data are listed for all the
Miras considered in this analysis in Table 3.
Examination of our $K-[12]$ data for R Dra indicates a spurious value,
the source of which was unclear.  This star was dropped from later
consideration in the analysis.

The observed Miras were separated into two sets: those with small
values of $K-[12]$ excess, between 0 and 1.25, which will be
indicative of Miras with lower mass loss rates, and those with
$K-[12]$ excess $> 1.75$, indicative of the higher mass loss rate
Miras; stars in the intermediate $K-[12]$ range were not considered
for this specific analysis (although were a part of analysis
associated with Figure 2, later in this section).
A plot of this is seen in Figure 1, where the stars with
low mass loss show systematically smaller sizes than those with
the more substantial mass loss; stars with multiple entries in
Table 3 are represented by a corresponding number of data points
in Figure 1. Fitting a line to each of the two data sets, we find
that $R=(1.01\pm0.50)\times P + (10\pm170)$ $R_\odot$ for those
Miras with little $K-[12]$ excess (line fit $\chi^2_\nu=0.41$),
and $R=(0.63\pm0.64)\times P + (270\pm245)$ $R_\odot$ for the
Miras with $K-[12]$ excess $> 1.75$ ($\chi^2_\nu=0.61$). Averaging
over periods of 300 to 450 days, we find that the higher mass loss
Miras appear on average 120 $R_\odot$ (roughly 30\%) larger than
their lower mass loss counterparts across all periods, with this
difference ranging from 145 $R_\odot$ to 90 $R_\odot$ as the
period increases from P=275$^d$ to 450$^d$. This correlation of
increased radius with $K-[12]$ excess is a clear indication that
the sizes we derive from our interferometric observations
progressively overestimate the Rosseland mean diameters for these
objects as mass loss increases. Noting the discussion in \S
\ref{linearRadius}, this result is independent of pulsation mode
assumptions.

Futhermore, the direction of that progression
appears to indicate that our diameters are systematically
biased towards larger sizes, which is consistent with the effects of
a circumstellar shell upon the interferometric observables.
The presence of
circumstellar structures substantial enough to be seen in our interferometric data
will grossly invalidate our assumptions regarding interpretation of the visibility
data in terms of a scaled uniform disk size, and tend to systematically
increase the derived stellar disk size at all spatial frequencies \citep{be01, sc01}.

Once we have established $K-[12]$ color excess dependency for the
Mira radii in our sample, we may inspect it to see if there is a
correlation between $K-[12]$ excess and inferred radius, using the
inferred radius as a function of period to then inspect the
indicated pulsation mode. As discussed in \S 3.5, the selected
period-luminosity relationship has little bearing upon the
resultant sizes, and as such, any assumption of pulsation mode
inherent in the referenced period-luminosity relationships do not
appear to alter our results. For this purpose, we utilized the
pulsation mode-dependent radius-period relationships as found in
\citet{os86}:
\begin{eqnarray}
1.86 \log R = 0.73 \log M +1.92+\log P & \textrm{ for Fundamental}\\
1.59 \log R = 0.51 \log M +1.60+\log P & \textrm{ for First overtone}
\end{eqnarray}
Additionally, for the purposes of using this relationship, we
shall adopt period-dependent masses, as illustrated in Figure 5
of \citet{wi00}, with a range of 15\% in mass versus period for $P\leq400$ days
(increasing from 0.7 to 1 $M_\odot$ with period),
and a range of 25\% in mass versus period for $P>400$ days (1 to 4 $M_\odot$).
Although this use of
a relationship derived from linear, rather than non-linear, modelling
might be questionable \citep{ba98,ya99}, it shall be sufficient to establish
the dependence of the measured radii upon mass loss rates.
Having established these theoretical mode-dependent regions, we may inspect
our data in reference to the fundamental and first overtone regions,
also seen in Figure 1.
There are two striking characteristics
to this aspect of the figure.
First, the tendency for the
Miras with large $K-[12]$ color excesses to lie in the first
overtone region.
Second, all of the stars lie above the region defined by
the fundamental overtone lines.
Both of these aspects are consistent with apparent size overestimation
due to circumstellar shell contamination of the visibility data.

Taking the sizes from the \citet{os86} fundamental mode
relationship and establishing predicted sizes for all of the stars in our
data set, we may plot the observed-to-predicted size ratio
versus the $K-[12]$ excess numbers for all of the Miras in our data set,
as seen in Figure 2.  There is a clear progression from smaller to
larger positive residuals as the excess increases, with predicted
scaling being in excess of 1.0 for all observed stars (excepting R Dra, as noted
above).
A similar exercise may be carried out in light of the first overtone
predictions, but this results in a scaling factor of less than 1.0 for stars
with $K-[12]$ excess less than 1.50, which is more than
half of our data set.  Scalings less than unity are inconsistent
with size enhancement due to circumstellar absorption. It is possible other physical
phenomena could appear to decrease
the size of the stars when viewed by the interferometer
(eg. bright spots), but this explanation is unlikely for two reasons.  First,
structures at spatial scales smaller than the gross size of the star will
need to satisfy a fortuitous geometry to properly affect our visibility
data.  While this is possible for specific examples in our data set, it seems
doubtful that this would present itself as a general effect across our
entire ensemble of data.
Second, it is unlikely that such structures will have sufficient contrast
at 2.2 $\mu$m to significantly affect the interferometric observables - for a 10\%
increase in size, a spot would have to be 33\% the diameter of the star,
properly aligned, and 1000K hotter than the surrounding photosphere.  Both the size and
temperature differential of such a spot do not seem credible.

Adjusting the angular diameters by this $K-[12]$ excess-dependent
scaling factor results in the average inferred effective temperature
rising from 2500K to 3200K, which is more consistent with
the expected effective temperature value for these stars at the Rosseland
mean radiating surface when they are pulsating in the fundamental mode
(HSW98, \citet{wi00}).

Measurements of the angular size of Mira at a variety of wavelengths
by other groups appears to support this hypothesis.  At 2.2 $\mu$m,
the $36.1 \pm 1.4$ mas size measured by \citet{ri92} is in agreement with a similar recent
measurement of $31.6$ mas at 2.26 $\mu$m by \citet{tu99b} (no error or
phase was given in this promising but preliminary result).
However, at 11.15 $\mu$m, \citet{we00} find an angular size of $47.8 \pm 0.5$
mas.  This larger size is consistent with the presence of
a circumstellar dust shell, which is in turn consistent with
the $K-[12]$ excess value that we compute to be in excess of
2 magnitudes (see Table 3).
Furthermore, \citet{tu99b} also measured the angular size of Mira at a variety
of other wavelengths: $\theta(1.24 \mu$m$)=23.3$ mas,
$\theta(1.65 \mu$m$)=28.3$ mas, and
$\theta(3.08 \mu$m$)=59.9$ mas.  The increased K band size
relative to the J band size appears to be consistent with our
expectation that the K band sizes are being enhanced
by the presence of circumstellar material.

\section{Conclusions}

The IOTA program of observing Mira variable stars has resulted in
a substantial increase in the number of near-infrared angular size
measurements for Miras; previous results were available for
$\approx25$ of these stars.  Determinations of $T_{EFF}$ and $R$
are possible, with the caveat that the derived Rosseland mean
radius values are potentially overestimated due to circumstellar
shells. The atmospheric models that are utilized in reducing the
visibility data clearly play a significant role in the results
obtained. Our data are consistent with at least some of the
observed Mira variables pulsating in the fundamental mode.
For those Miras that appear to be too large to be pulsating in the
fundamental mode, there appears to be a correlation between K-[12]
color excess and size, reflective of mass loss
enhancing the apparent size.
If our inference that molecular absorption about the Miras is
responsible for an increased apparent K band size of these
objects, a number of effects should be observable, the most
noticable of which will be the size of the Miras will appear to be
smaller in the J and H bands than in the K band, where such
absorption is less. The preliminary results of \citet{tu99b}
appear to support this conclusion.  We predict that an angular
size pulsation mode study of these objects carried out at 1.6
$\mu$m or particularly
1.2 $\mu$m will result in unequivocal evidence for fundamental
mode pulsation for most if not all Miras.



\acknowledgments

We would like to thank the staff at the Center for Astrophysics
for a generous allotment of telescope time so that this project
could be carried out, and Ron Canterna for liberal use of computer
resources for data reduction.  We acknowledge particularly fruitful discussions
with Michael Scholz and Lee Anne Willson.
This research has been partially supported by NSF
grant AST-9021181 to the University of Wyoming, the Wyoming Space
Grant Consortium, and NASA Grant NGT-40050. This research has made
use of the SIMBAD database and the AFOEV database, both operated
by the CDS, Strasbourg, France.  In this research, we have used,
and acknowledge with thanks, data from the AAVSO International
Database, based on observations submitted to the AAVSO by variable
star observers worldwide.  Portions of this work were performed at
the Jet Propulsion Laboratory, California Institute of Technology
under contract with the National Aeronautics and Space
Administration.

\input{vanBelle.tab1}

\input{vanBelle.tab2}
\input{vanBelle.tab3}

\clearpage

\begin{figure}
     \epsscale{1.0}
     \plotone{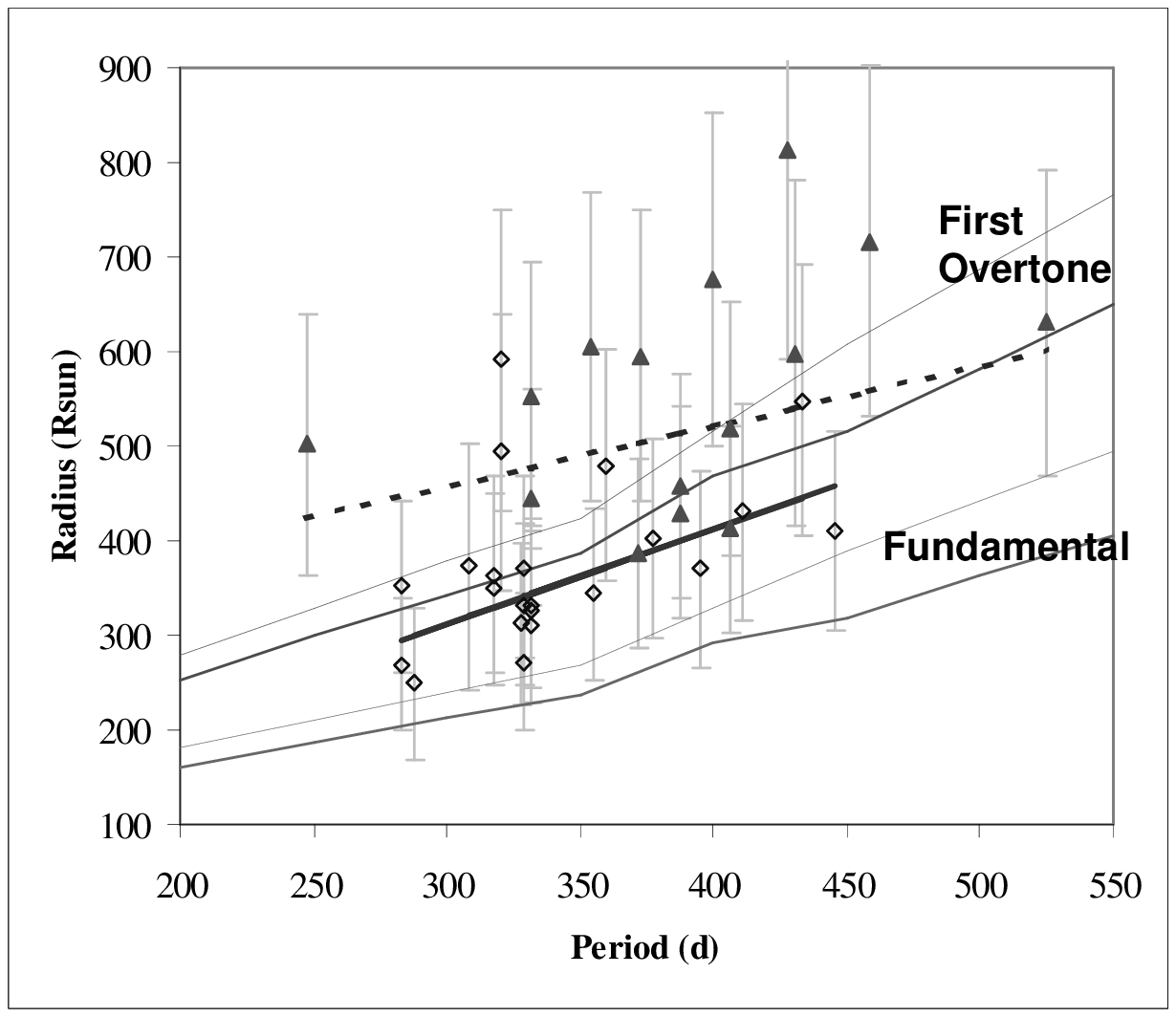}
\caption{Radius versus period for the Miras observed.  Open
diamonds are for Miras with $K-[12]$ excess between +0.0 and
+1.25, indicating less dusty mass loss, and filled triangles are
for Miras with $K-[12]$ excess $>$ +1.75, indicating more lass
loss. Comparing fit lines for the lower and higher mass loss Miras
(solid and dotted, respectively), the Miras with greater dusty
mass loss appear on average 120 $R_\odot$ larger. The fundamental
and first overtone regions are derived from the radius-period
relationships found in \citet{os86}.  See \S \ref{discussion} for
details of the line fits.} \label{RvsP}
\end{figure}

\begin{figure}
     \epsscale{1.0}
     \plotone{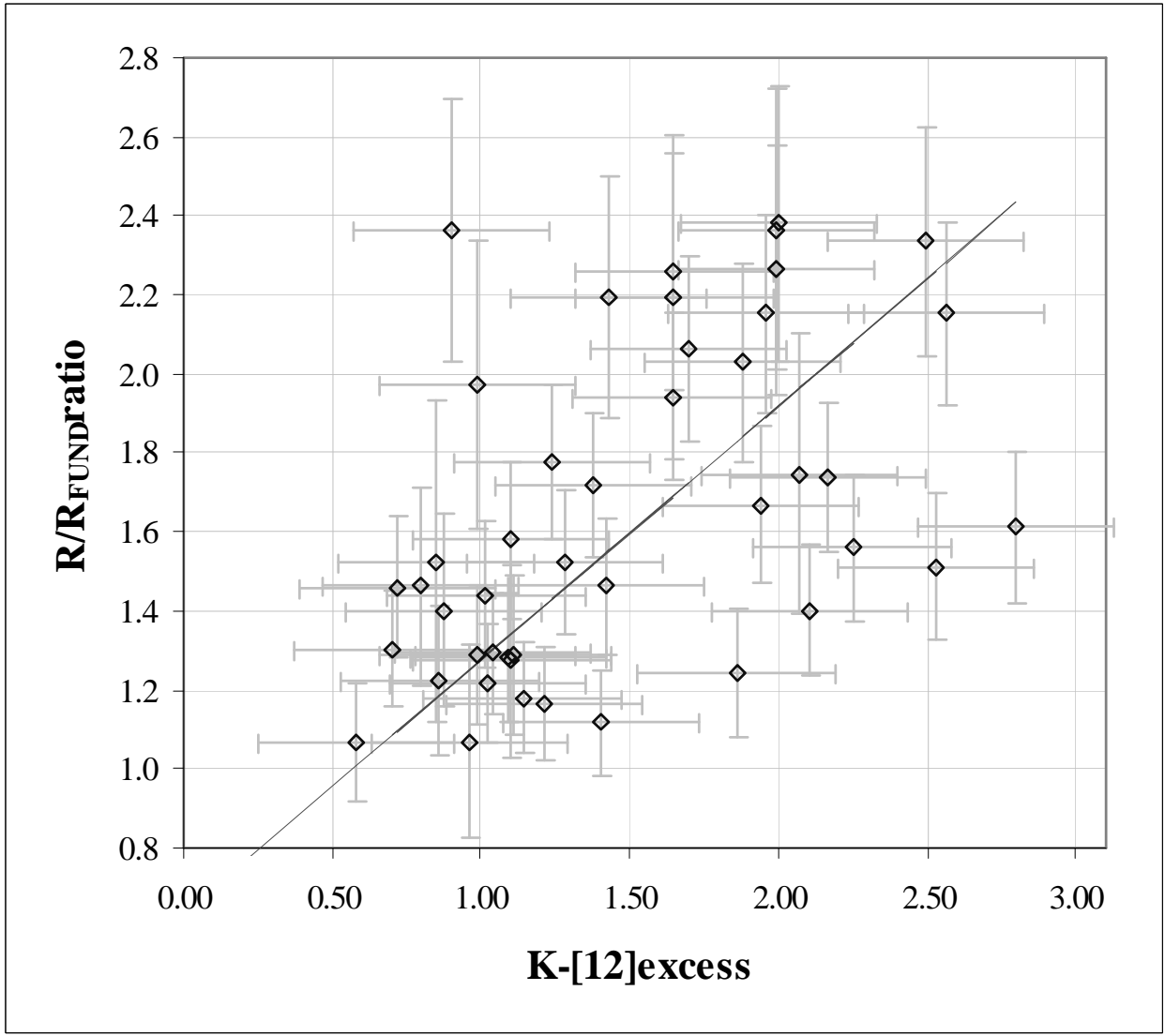}
\caption{Ratio of observed radius to predicted fundamental mode
radius versus $K-[12]$ excess.}
\label{RvsPres}
\end{figure}

\end{document}

%% file: vanBelle.tab1.tex
\begin{deluxetable}{ccccccccc}
\tablecolumns{9}
\tabletypesize{\scriptsize}
\tablewidth{0pc}
\tablecaption{The observed data.}
\tablehead{
\colhead{} & \colhead{Period}   & \colhead{$\phi=0$}    & \colhead{Obs.} &
\colhead{Obs.}    & \colhead{Phase}   & \colhead{$B_P$}    & \colhead{}
& \colhead{$\theta_{UD}$}\\
\colhead{Star} & \colhead{$P$}   & \colhead{JD}    &
\colhead{Date} & \colhead{JD}    & \colhead{$\phi$}   &
\colhead{(m)}    & \colhead{$V$\tablenotemark{a}} & \colhead{(mas)} } \startdata
RR Aql & 399.8 & 50,046 & 96-Jun-04 & 50239 & 0.48 & 31.13 & 0.465 & $10.73 \pm 0.66$ \\
RT Aql & 328.6 & 49,875 & 96-Jun-03 & 50238 & 0.10 & 37.00 & 0.705 & $7.24 \pm 0.42$ \\
 & & & 96-Jun-07 & 50242 & 0.12 & 35.45 & 0.582 & $$ \\
V Cam & 524.5 & 50,190 & 96-Oct-09 & 50366 & 0.34 & 26.57 & 0.759 & $8.36 \pm 0.40$ \\
 & & & 96-Oct-09 & 50366 & 0.34 & 27.41 & 0.740 & $$ \\
 & && 96-Oct-09 & 50366 & 0.34 & 28.04 & 0.663 & $$ \\
 & & & 96-Oct-09 & 50366 & 0.34 & 28.64 & 0.692 & $$ \\
V Cas & 226.6 & 50,289 & 96-Oct-06 & 50363 & 0.33 & 34.69 & 0.740 & $6.30 \pm 0.66$ \\
Y Cas & 410.7 & 50,303 & 96-Oct-07 & 50364 & 0.15 & 35.95 & 0.643 & $7.28 \pm 0.59$ \\
S CrB & 359.5 & 49,725 & 96-Mar-07 & 50150 & 0.18 & 21.21 & 0.754 & $11.35 \pm 0.26$ \\
 & & & 96-Mar-07 & 50150 & 0.18 & 21.21 & 0.662 & $$ \\
 & & & 96-Mar-07 & 50150 & 0.18 & 21.21 & 0.726 & $$ \\
 & & & 96-Mar-08 & 50151 & 0.19 & 21.21 & 0.663 & $$ \\
 & & & 96-Mar-08 & 50151 & 0.19 & 21.21 & 0.641 & $$ \\
 & & & 96-Mar-08 & 50151 & 0.19 & 21.21 & 0.639 & $$ \\
 & & & 96-Mar-08 & 50151 & 0.19 & 21.21 & 0.619 & $$ \\
 & & & 96-Mar-08 & 50151 & 0.19 & 21.21 & 0.611 & $$ \\
 & & & 96-Mar-08 & 50151 & 0.19 & 21.21 & 0.555 & $$ \\
 & & & 96-Mar-08 & 50151 & 0.19 & 21.20 & 0.598 & $$ \\
 & & & 96-Mar-12 & 50155 & 0.20 & 38.22 & 0.357 & $$ \\
 & & & 96-Mar-12 & 50155 & 0.20 & 38.21 & 0.368 & $$ \\
R CVn & 329.2 & 50,150 & 96-Mar-13 & 50156 & 0.02 & 37.00 & 0.681 & $6.63 \pm 0.59$ \\
 & & & 96-May-29 & 50233 & 0.25 & 37.60 & 0.583 & $7.66 \pm 0.28$ \\
 & & & 96-May-30 & 50234 & 0.26 & 37.57 & 0.590 & $$ \\
 & & & 96-May-30 & 50234 & 0.26 & 37.66 & 0.599 & $$ \\
 & & & 96-Jun-07 & 50242 & 0.28 & 35.52 & 0.581 & $$ \\
BG Cyg & 287.7 & 49,524 & 96-Jun-07 & 50242 & 0.50 & 35.45 & 0.877 & $4.14 \pm 0.83$ \\
DG Cyg & 463.9 & 49,891 & 96-May-29 & 50233 & 0.74 & 36.80 & 0.784 & $5.36 \pm 0.66$ \\
R Dra & 246.0 & 50,431 & 97-Jul-05 & 50635 & 0.83 & 20.90 & 0.658 & $12.04 \pm 1.36$ \\
 & & & 97-Jul-05 & 50635 & 0.83 & 20.57 & 0.692 & $$ \\
RU Her & 485.6 & 49,800 & 96-Jun-02 & 50237 & 0.90 & 37.58 & 0.507 & $8.71 \pm 0.39$ \\
 & & & 96-Jun-03 & 50238 & 0.90 & 37.57 & 0.457 & $$ \\
U Her & 406.4 & 49,880 & 96-Jun-02 & 50237 & 0.88 & 37.31 & 0.263 & $11.18 \pm 0.60$ \\
R LMi & 372.8 & 49,600 & 96-Mar-07 & 50150 & 0.48 & 21.19 & 0.534 & $14.40 \pm 0.66$ \\
 & & & 96-Mar-07 & 50150 & 0.48 & 21.19 & 0.668 & $$ \\
 & & & 96-Mar-07 & 50150 & 0.48 & 21.07 & 0.529 & $$ \\
 & & & 96-Mar-07 & 50150 & 0.48 & 21.05 & 0.552 & $$ \\
RT Oph & 427.4 & 49,925 & 96-Jun-07 & 50242 & 0.74 & 35.16 & 0.717 & $6.52 \pm 0.64$ \\
UU Peg & 458.5 & 49,690 & 96-Jun-03 & 50238 & 0.20 & 36.96 & 0.417 & $9.56 \pm 0.56$ \\
Z Peg & 331.4 & 50,711 & 96-Oct-04 & 50361 & 0.94 & 37.64 & 0.838 & $4.50 \pm 0.71$ \\
RR Per & 390.1 & 50,039 & 96-Oct-07 & 50364 & 0.83 & 35.99 & 0.679 & $6.83 \pm 0.61$ \\
U Per & 317.5 & 50,165 & 96-Oct-06 & 50363 & 0.62 & 33.93 & 0.811 & $5.41 \pm 0.75$ \\
BG Ser & 394.8 & 48,450 & 96-Jun-04 & 50239 & 0.53 & 30.51 & 0.769 & $6.71 \pm 0.78$ \\
S Ser & 375.9 & 49,875 & 96-Mar-11 & 50154 & 0.74 & 36.98 & 0.733 & $5.51 \pm 0.42$ \\
 & & & 96-Mar-11 & 50154 & 0.74 & 36.89 & 0.796 & $$ \\
 & & & 96-Mar-11 & 50154 & 0.74 & 36.89 & 0.796 & $$ \\
 & & & 96-Jun-08 & 50243 & 0.98 & 32.26 & 0.832 & $$ $5.35 \pm 0.82$\\
S UMi & 328.2 & 49,850 & 96-Jun-06 & 50241 & 0.19 & 26.67 & 0.752 & $7.98 \pm 0.87$ \\
RS Vir & 354.1 & 49,325 & 96-Jun-10 & 50245 & 0.60 & 32.38 & 0.683 & $7.54 \pm 0.67$ \\
\hline
R Cas\tablenotemark{b} & 430.8 & 50,940 & 95-Oct-4  & 49995 & 0.81 & 35.27 & 0.125 & ${22.03}^{+2.13}_{-4.13}$\\
       &&& 95-Oct-5 & 49996 & 0.81 & 36.30 & 0.1273 &\\
\enddata
\tablenotetext{a}{Standard nightly error is $\Delta V=0.0509$.}
\tablenotetext{b}{R Cas data is reanalysis of visibility data from van Belle et al. (1996).}
\end{deluxetable}

%% file: vanBelle.tab2.tex
\begin{deluxetable}{cccccccc}
\tablecolumns{8}
\tabletypesize{\scriptsize}
\tablewidth{0pc}
\tablecaption{Derived parameters: bolometric flux, Rosseland angular diameter, effective temperature, distance and linear radius.}
\tablehead{
\colhead{} & \colhead{}   & \colhead{}    & \colhead{} & \colhead{} &
\colhead{$F_{BOL}$}    & \colhead{$\theta_R$}   & \colhead{$T_{EFF}$}\\
\colhead{Star} & \colhead{Date}   & \colhead{$V$}    &
\colhead{$K$} & \colhead{$V-K$} & \colhead{$10^{-8}$ erg cm$^{-2}$s$^{-1}$}
& \colhead{(mas)}   & \colhead{(K)} } \startdata
RR Aql & 96-Jun-04 & 12.8 & $0.65 \pm 0.11$ & $12.1 \pm 0.5$ & $78.4 \pm 11.8$ & $10.73 \pm 0.79$ & $2127 \pm 111$\\
RT Aql & 96-Jun-03 & 9.5 & $0.50 \pm 0.12$ & $9.0 \pm 0.5$ & $101.7 \pm 15.2$ & $7.24 \pm 0.51$ & $2763 \pm 125$\\
V Cam & 96-Oct-09 & 13.0 & $0.63 \pm 0.10$ & $12.4 \pm 0.5$ & $79.1 \pm 11.9$ & $8.36 \pm 0.52$ & $2414 \pm 86$\\
V Cas & 96-Oct-06 & 10.8 & $1.88 \pm 0.20$ & $8.9 \pm 0.6$ & $28.6 \pm 4.3$ & $6.30 \pm 0.71$ & $2157 \pm 146$\\
Y Cas & 96-Oct-07 & 11.2 & $0.47 \pm 0.20$ & $10.7 \pm 0.6$ & $97.4 \pm 14.6$ & $7.28 \pm 0.66$ & $2727 \pm 160$\\
S CrB & 96-Mar-07 & 10.4 & $-0.07 \pm 0.03$ & $10.5 \pm 0.4$ & $160.5 \pm 24.1$ & $11.35 \pm 0.52$ & $2473 \pm 46$\\
R CVn & 96-Mar-13 & 8.5 & $0.01 \pm 0.46$ & $8.5 \pm 0.8$ & $134.8 \pm 20.2$ & $7.66 \pm 0.41$ & $2882 \pm 90$\\
 & 96-May-29 & 10.5 & $0.14 \pm 0.05$ & $10.4 \pm 0.4$ & $162.9 \pm 24.4$ & $6.63 \pm 0.64$ & $3248 \pm 200$\\
BG Cyg & 96-Jun-07 & 11.5 & $1.05 \pm 0.12$ & $10.5 \pm 0.5$ & $57.9 \pm 8.7$ & $4.14 \pm 0.84$ & $3175 \pm 344$\\
DG Cyg & 96-May-29 & 11.0 & $1.18 \pm 0.20$ & $9.8 \pm 0.6$ & $52.6 \pm 7.9$ & $5.36 \pm 0.69$ & $2723 \pm 204$\\
R Dra & 97-Jul-05 & 11.3 & $0.66 \pm 0.14$ & $10.6 \pm 0.5$ & $82.6 \pm 12.4$ & $12.04 \pm 1.44$ & $2034 \pm 139$\\
RU Her & 96-Jun-02 & 11.3 & $0.25 \pm 0.12$ & $11.0 \pm 0.5$ & $114.7 \pm 17.2$ & $8.71 \pm 0.52$ & $2596 \pm 109$\\
U Her & 96-Jun-02 & 10.3 & $-0.04 \pm 0.13$ & $10.3 \pm 0.5$ & $158.9 \pm 23.8$ & $11.18 \pm 0.75$ & $2486 \pm 125$\\
R LMi & 96-Mar-07 & 13.0 & $-0.16 \pm 0.06$ & $13.2 \pm 0.4$ & $158.6 \pm 23.8$ & $14.40 \pm 0.87$ & $2189 \pm 87$\\
RT Oph & 96-Jun-07 & 14.0 & $2.03 \pm 0.50$ & $12.0 \pm 0.8$ & $22.1 \pm 3.3$ & $6.52 \pm 0.69$ & $1989 \pm 129$\\
UU Peg & 96-Jun-03 & 13.8 & $0.82 \pm 0.15$ & $12.9 \pm 0.5$ & $64.9 \pm 9.7$ & $9.56 \pm 0.68$ & $2149 \pm 111$\\
Z Peg & 96-Oct-04 & 9.3 & $1.24 \pm 0.20$ & $8.0 \pm 0.6$ & $53.7 \pm 8.0$ & $4.50 \pm 0.73$ & $2987 \pm 267$\\
RR Per & 96-Oct-07 & 13.1 & $1.50 \pm 0.20$ & $11.6 \pm 0.6$ & $36.5 \pm 5.5$ & $6.83 \pm 0.66$ & $2202 \pm 135$\\
U Per & 96-Oct-06 & 9.3 & $1.05 \pm 0.20$ & $8.2 \pm 0.6$ & $63.3 \pm 9.5$ & $5.41 \pm 0.78$ & $2839 \pm 230$\\
BG Ser & 96-Jun-04 & 11.9 & $0.38 \pm 0.19$ & $11.5 \pm 0.5$ & $103.1 \pm 15.5$ & $6.71 \pm 0.83$ & $2879 \pm 207$\\
S Ser & 96-Mar-11 & 11.8 & $2.03 \pm 0.26$ & $9.8 \pm 0.6$ & $26.9 \pm 4.0$ & $5.35 \pm 0.84$ & $2305 \pm 201$\\
 & 96-Jun-08 & 11.8 & $2.03 \pm 0.45$ & $9.8 \pm 0.7$ & $24.1 \pm 3.6$ & $5.51 \pm 0.48$ & $2209 \pm 106$\\
S UMi & 96-Jun-06 & 10.2 & $-0.09 \pm 0.15$ & $10.3 \pm 0.5$ & $166.4 \pm 25.0$ & $7.98 \pm 0.93$ & $2977 \pm 206$\\
RS Vir & 96-Jun-10 & 13.2 & $1.36 \pm 0.20$ & $11.8 \pm 0.6$ & $41.2 \pm 6.2$ & $7.54 \pm 0.74$ & $2160 \pm 133$\\
\hline
R Cas\tablenotemark{a} &  95-Oct-4  & 11.1 & $-1.29\pm 0.13 $& $12.4 \pm 0.4$ & $447.4 \pm 85.2$ & ${23.03}^{+2.13}_{-4.13}$ & ${2239}^{+152}_{-236}$\\
\enddata
\tablenotetext{a}{R Cas data is reanalysis of visibility data from van Belle et al. (1996).}
\end{deluxetable}

%% file: vanBelle.tab3.tex
\begin{deluxetable}{ccccccccc}
\tablecolumns{9}
\tabletypesize{\scriptsize}
\tablewidth{0pc}
\tablecaption{
Mira variable absolute K magnitudes, distances, radii, IRAS 12$\mu$m magnitudes, $K-[12]$ colors,
and estimated $K-[12]$ color excesses.}
\tablehead{
\colhead{} & \colhead{} & \colhead{$M_K$} & \colhead{dist} &
\colhead{Radius}    & \colhead{$[12]$}   & \colhead{$K-[12]$} &
\colhead{$K-[12]$} & \colhead{}\\
\colhead{Star} & \colhead{Date}   & \colhead{(mag)}    &
\colhead{(pc)} & \colhead{($R_\odot$)} & \colhead{(mag)} &
\colhead{(mag)}&
\colhead{excess (mag)} & \colhead{Ref}
}
\startdata
R Aql & 95-Jun-10 & -7.55 & 224 & $259 \pm 67$ & -2.88 & 2.04 & 1.13 & 1 \\
 & 95-Oct-06 & -7.55 & 224 & $328 \pm 85$ & -2.88 & 2.11 & 1.20 & 1 \\
RR Aql & 96-Jun-04 & -8.09 & 561 & $648 \pm 169$ & -2.67 & 3.32 & 2.41 & 2 \\
RT Aql & 96-Jun-05 & -7.46 & 392 & $305 \pm 79$ & -1.05 & 1.55 & 0.64 & 2 \\
R Aqr & 95-Jul-11 & -8.05 & 272 & $438 \pm 114$ & -4.37 & 3.36 & 2.45 & 1 \\
 & 95-Oct-07 & -8.05 & 272 & $412 \pm 106$ & -4.37 & 3.63 & 2.72 & 1 \\
U Ari & 77-Sep-03 & -7.98 & 779 & $801 \pm 205$ & -0.99 & 2.47 & 1.56 & 3 \\
R Aur & 95-Oct-03 & -8.46 & 342 & $407 \pm 105$ & -3.03 & 2.24 & 1.33 & 1 \\
V Cam & 96-Oct-09 & -8.87 & 796 & $716 \pm 185$ & -2.14 & 2.77 & 1.86 & 2 \\
R Cas & 95-Oct-04 & -8.28 & 250 & $593 \pm 181$ & -4.19 & 2.90 & 1.99 & 2 \\
T Cas & 95-Oct-04 & -8.38 & 302 & $416 \pm 108$ & -2.95 & 1.97 & 1.06 & 1 \\
V Cas & 96-Oct-06 & -7.33 & 696 & $472 \pm 129$ & -0.94 & 2.82 & 1.91 & 2 \\
Y Cas & 96-Oct-07 & -8.14 & 527 & $413 \pm 110$ & -1.34 & 1.82 & 0.91 & 2 \\
$o$ Cet & 90-Aug-21 & -7.80 & 121 & $482 \pm 124$ & -5.59 & 3.39 & 2.48 & 4 \\
 & 90-Sep-19 & -7.80 & 121 & $468 \pm 120$ & -5.59 & 2.99 & 2.08 & 4 \\
S CrB & 96-Mar-08 & -7.93 & 375 & $457 \pm 116$ & -2.13 & 2.07 & 1.16 & 2 \\
R CVn & 96-Mar-13 & -7.79 & 391 & $279 \pm 75$ & -1.40 & 1.41 & 0.50 & 2 \\
 & 96-May-31 & -7.79 & 391 & $322 \pm 82$ & -1.40 & 1.74 & 0.83 & 2 \\
BG Cyg & 96-Jun-07 & -7.57 & 530 & $236 \pm 76$ & -0.74 & 1.79 & 0.88 & 2 \\
DG Cyg & 96-May-29 & -8.50 & 865 & $499 \pm 140$ & ... & ... & ... & 2 \\
R Dra & 97-Jul-05 & -7.32 & 395 & $511 \pm 142$ & 0.42 & 0.24 & -0.67 & 2 \\
RU Her & 96-Jun-02 & -8.64 & 610 & $572 \pm 147$ & -1.97 & 2.25 & 1.34 & 2 \\
S Her & 95-Jul-07 & -7.68 & 677 & $354 \pm 124$ & -0.21 & 1.68 & 0.77 & 1 \\
U Her & 95-Jun-10 & -8.11 & 377 & $431 \pm 114$ & -3.12 & 2.69 & 1.78 & 1 \\
 & 96-Jun-02 & -8.11 & 377 & $453 \pm 117$ & -3.12 & 3.08 & 2.17 & 2 \\
R Leo & 90-May-02 & -7.81 & 115 & $409 \pm 105$ & -4.71 & 2.21 & 1.30 & 5 \\
R LMi & 96-Mar-07 & -7.98 & 367 & $569 \pm 146$ & -2.94 & 2.78 & 1.87 & 2 \\
RT Oph & 96-Jun-07 & -8.26 & 1,142 & $801 \pm 217$ & -0.80 & 2.83 & 1.92 & 2 \\
X Oph & 95-Jul-07 & -7.80 & 225 & $314 \pm 81$ & -2.90 & 1.87 & 0.96 & 1 \\
 & 95-Oct-07 & -7.80 & 225 & $298 \pm 78$ & -2.90 & 1.85 & 0.94 & 1 \\
S Ori & 95-Jul-07 & -8.30 & 481 & $545 \pm 142$ & -1.82 & 1.93 & 1.02 & 1 \\
U Ori & 95-Oct-08 & -7.98 & 310 & $370 \pm 96$ & -3.45 & 2.93 & 2.02 & 1 \\
R Peg & 95-Jul-07 & -8.00 & 435 & $476 \pm 122$ & -2.03 & 2.53 & 1.62 & 1 \\
 & 95-Oct-06 & -8.00 & 435 & $455 \pm 119$ & -2.03 & 1.85 & 0.94 & 1 \\
S Peg & 95-Oct-06 & -7.74 & 675 & $459 \pm 135$ & -0.37 & 1.82 & 0.91 & 1 \\
 & 95-Jul-08 & -7.74 & 675 & $574 \pm 154$ & -0.37 & 1.73 & 0.82 & 1 \\
UU Peg & 96-Jun-03 & -8.47 & 721 & $742 \pm 193$ & -1.89 & 2.71 & 1.80 & 2 \\
Z Peg & 96-Oct-04 & -7.80 & 640 & $310 \pm 92$ & -0.70 & 1.93 & 1.02 & 2 \\
RR Per & 96-Oct-07 & -8.06 & 816 & $599 \pm 161$ & -0.76 & 2.26 & 1.35 & 2 \\
U Per & 95-Oct-04 & -7.73 & 559 & $352 \pm 101$ & -0.66 & 1.63 & 0.72 & 1 \\
 & 96-Oct-06 & -7.73 & 559 & $325 \pm 94$ & -0.66 & 1.71 & 0.80 & 2 \\
BG Ser & 96-Jun-04 & -8.07 & 491 & $354 \pm 99$ & -1.56 & 1.94 & 1.03 & 2 \\
R Ser & 95-Jul-07 & -7.91 & 356 & $328 \pm 86$ & -2.07 & 1.92 & 1.01 & 1 \\
S Ser & 96-Mar-11 & -8.00 & 1,012 & $601 \pm 159$ & -0.45 & 2.48 & 1.57 & 2 \\
 & 96-Jun-08 & -8.00 & 1,012 & $583 \pm 172$ & -0.45 & 2.48 & 1.57 & 2 \\
S UMi & 96-Jun-06 & -7.78 & 345 & $296 \pm 82$ & -1.78 & 1.69 & 0.78 & 2 \\
RS Vir & 96-Jun-10 & -7.90 & 712 & $577 \pm 155$ & -1.46 & 2.82 & 1.91 & 2 \\
\enddata
\tablerefs{1. \citet{va96},  2. van Belle et al. (2002),  3. \citet{ri79},
4. \citet{ri92},  5. \citet{di91}.}
\end{deluxetable}
